\documentclass{emulateapj}

\slugcomment{Draft of \today}

\shorttitle{An Apparent Precessing Helical Outflow from a Massive Evolved Star: Evidence for Binary Interaction}
\shortauthors{Lau et al.}

\usepackage{amsmath}
\usepackage{graphicx}
\usepackage[section] {placeins}
\usepackage{hyperref}
\usepackage{ulem}

\newcommand{\beq}{\begin{equation}}
\newcommand{\eeq}{\end{equation}}

\begin{document}

\title{An Apparent Precessing Helical Outflow from a Massive Evolved Star: Evidence for Binary Interaction}

\author{R. M. Lau\altaffilmark{1,2},
M. J. Hankins\altaffilmark{1},
T. L. Herter\altaffilmark{1},
M. R. Morris\altaffilmark{3},
E. A. C. Mills\altaffilmark{4},
M. E. Ressler\altaffilmark{2},
}
\altaffiltext{1}{Astronomy Department, Cornell University, Ithaca, NY 14853-6801, USA}
\altaffiltext{2}{Jet Propulsion Laboratory, California Institute of Technology, 4800 Oak Grove Drive, Pasadena, CA 91109, USA}
\altaffiltext{3}{Department of Physics and Astronomy, University of California, Los Angeles, 430 Portola Plaza, Los Angeles, CA 90095, USA}
\altaffiltext{4}{National Radio Astronomy Observatory, P.O. Box O 1009, Lopezville Drive, Socorro, NM 87801, USA}

\begin{abstract}

Massive, evolved stars play a crucial role in the metal-enrichment, dust budget, and energetics of the interstellar medium; however, the details of their evolution are uncertain because of their rarity and short lifetimes before exploding as supernovae. Discrepancies between theoretical predictions from single-star evolutionary models and observations of massive stars have evoked a shifting paradigm that implicates the importance of binary interaction. We present mid- to far-infrared observations from the Stratospheric Observatory for Infrared Astronomy (SOFIA) of a conical ``helix'' of warm dust ($\sim180$ K) that appears to extend from the Wolf-Rayet star WR102c. Our interpretation of the helix is a precessing, collimated outflow that emerged from WR102c during a previous evolutionary phase as a rapidly rotating luminous blue variable. We attribute the precession of WR102c to gravitational interactions with an unseen compact binary companion whose orbital period can be constrained to $800\,\mathrm{d}<P<1400$ d from the inferred precession period, $\tau_p\sim1.4\times10^4$ yr, and limits imposed on the stellar and orbital parameters of the system. Our results concur with the range of orbital periods ($P\lesssim1500$ d) where spin-up via mass exchange is expected to occur for massive binary systems.

\end{abstract}

\maketitle

\section{Introduction}
Massive stars that are born with an initial mass greater than $\sim20 \,\mathrm{M}_\odot$ have a profound influence on the interstellar medium (ISM) of their host galaxies. Such massive stars are dominant sources of optical and ultraviolet (UV) photons and are responsible for heating nearby dust that is, in turn, used as a probe for measuring star formation in distant galaxies of the early Universe (Kennicutt 1998). The extreme luminosity and high effective temperature of these massive stars can produce fast winds from their surfaces driven by radiation pressure (Kudritzki \& Puls 2000). This mass loss will directly influence the evolution and death of massive stars since a star's mass will dictate its power output, effective temperature, and nucleosynthetic products (Smith 2014). In the final phases of their lives after leaving the main sequence, massive stars can undergo episodes of enhanced, violent mass loss (Humphreys \& Davidson 1994) that impacts the surrounding ISM (Freyer, Hensler, \& Yorke 2003). The resulting supernova explosion can drive powerful shocks into the surrounding medium as well as enrich its metal content, thereby influencing the chemical evolution and dust budget of the host galaxy (Gall et al. 2011 and ref. therein). Massive stars that exhibit rapid rotation are also thought to be progenitors of gamma ray bursts (GRBs); these explosions are the most luminous transient events ever and are powerful probes of star and galaxy formation throughout the early Universe (Woosley \& Bloom 2006; Gehrels, Ramirez-Ruiz, \& Fox 2009 and ref. therein).

Until recently, massive stars were generally thought to evolve as single-star systems, and their mass loss was attributed to steady winds that implied smooth scaling relations between the mass loss rate and changes in luminosity, temperature, and metallicity (e.g. Nieuwenhuijzen \& de Jager 1990; Nugis \& Lamers 2000). Single-star evolutionary models, however, failed to reproduce the prevalence of observed phases of episodic and extreme mass loss in post-main sequence stars as well as their highly spun-up rotation velocities (Groh, Hillier, Damineli 2006; Groh et al. 2009). Notably, the large observed fraction of core-collapse supernovae lacking hydrogen (e.g. Type Ibc and IIb) cannot be explained by single-star evolutionary models for a standard initial mass function, which suggests that the hydrogen envelopes were stripped from their progenitors due to binary Roche-lobe overflow (Smith et al. 2011). These issues have provoked a paradigm shift in the understanding of mass loss and evolution of massive stars: binarity is recognized to have a dominant influence in their evolution since observations now suggest over $\sim70\%$ of massive stars are expected to exchange mass or merge with a companion over their lifetime (Sana et al. 2012). Mass exchange is predicted to occur for binary systems with orbital periods less than $\sim1500$ days (Podsiadlowski, Joss, \& Hsu 1992; Sana et al. 2012; de Mink et al. 2013). The mass-gaining companion also gains angular momentum and spins up the star, which leads to an asymmetry in its winds (Dwarkadas \& Owocki 2002) and can affect its internal composition by triggering mixing (Brott \emph{et al}. 2011). The evolutionary track of both mass gainer and donor are therefore drastically different from their alternate evolutionary path as a single star. However, since massive, evolved stars are quite rare due to their short lifetimes, there is a paucity of observations from which the impact of binarity on their mass-loss and evolution can be determined.

\begin{figure*}[t]
	\centerline{\includegraphics[scale=.38]{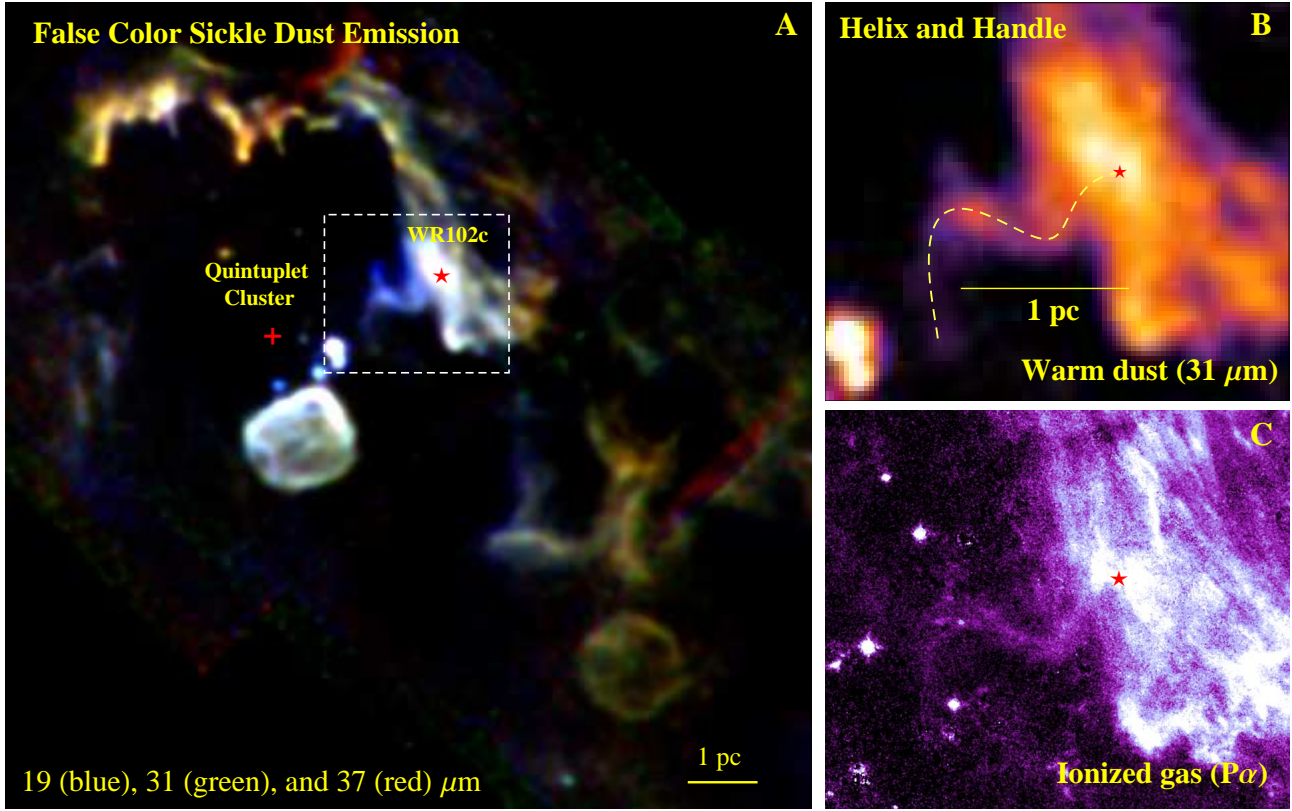}}
	\caption{(A) False color image of the Sickle, the illuminated inner edge of a dense molecular cloud heated by the Quintuplet cluster overlaid with a white dashed box indicating the location of the ``helix" extending from WR102c. The cross and star correspond to the approximate, projected center of the Quintuplet cluster and WR102c, respectively. (B) Zoomed 31 $\mu$m image of the helix. The overlaid dashed line traces the curvature of the helix. (C) Zoomed Paschen-$\alpha$ (Wang et al. 2010; Dong et al. 2011) image of the same region as (B). North is up and east is to the left.}
	\label{fig:SickleHelix}
\end{figure*}

Imaging observations from the Stratospheric Observatory for Infrared Astronomy (SOFIA) taken in the mid to far-infrared (IR) reveal a dusty $\sim1.5$ pc long, conical ``helix"-shaped trail extending from the massive, evolved star, WR102c (Figer, McLean, \& Morris 1999; Barniske, Oskinova, \& Hamann 2008, Steinke et al. 2015; Fig.~\ref{fig:SickleHelix}A-C). WR102c is a Wolf-Rayet (WR) star, a descendant of a massive O star in one of the final evolutionary phases before exploding as a supernova (Crowther 2007 and ref. therein). Near-infrared (IR) spectroscopy of WR102c reveal the presence of helium and nitrogen emission lines with no significant contribution from hydrogen (Figer, McLean, \& Morris 1999, Steinke et al. 2015), which classifies WR102c as a hydrogen-poor/free, nitrogen-rich WR (WN) subtype. WR102c is located in the vicinity of the ``handle" of the nearby ``Sickle" HII region (Yusef-Zadeh \& Morris 1987; Fig.~\ref{fig:SickleHelix}a) and is $\sim3$ pc in projection from the massive, young (4 - 6 Myr) Quintuplet star cluster near the Galactic center (Okuda et al. 1990; Figer, McLean, \& Morris 1999; Liermann, Hamann, \& Oskinova 2009). The Quintuplet cluster contains a population of massive, evolved stars, which suggests WR102c is a cluster member despite its relative isolation. Barniske, Oskinova, \& Hamann (2008) claimed that WR102c is too young to have formed in the Quintuplet cluster due to the high luminosity and mass inferred from models fit to IR photometry; however, they mistakenly adopted the near-IR flux from a bright, nearby star. Recent work has shown that evolutionary tracks and models fit to the near-IR spectra of WR102c (Steinke et al. 2015) imply an age $\lesssim 6$ Myr, consistent with the age of the Quintuplet cluster. 

We claim that the helix is associated with WR102c based on three results from our observations: 1) The dust and gas morphology and orientation of the helix are consistent with originating as an outflow of WR102c. 2) Observed dust temperatures and model fits to spectral energy densities of the region reveal that the helix must be heated by a nearby, luminous source such as WR102c and be composed of very small grains, whereas the surrounding dust in the Sickle is composed of larger grains. 3) There is a lack of cold dust emission at sub-millimeter wavelengths from the helix, unlike the Sickle, which indicates that the helix is not associated with a dense molecular cloud along the line of sight. Dozens of WR stars exhibit dusty, circumstellar nebulae, some of which show a bipolar morphology (Toal\'{a} et al. 2015). WR nebulae are interpreted as interactions between fast ($\sim1000$ km/s, Crowther 2007) WR winds and the ejecta from a previous red supergiant (RSG) or luminous blue variable (LBV, Conti 1984; Humphreys \& Davidson 1994) phase, where the star underwent enhanced mass loss through a slow ($\sim10-100$ km/s) and dense outflow. A dusty helical ``jet" and bipolar outflows have also been previously reported around the LBV AG Carinae (Paresce \& Nota 1989, Nota et al. 1992, 1995).


We posit that the WR102c helix was produced by a highly collimated, precessing outflow during a prior LBV phase. Our claim is substantiated by radial velocity follow-up observations in the near-infrared using Palomar/TripleSpec. Radial velocity shifts of the Brackett-$\gamma$ ($\lambda=2.16\mu$m) along the helix indicate outflow velocities that are consistent with the expansion speed of LBV nebulae ($\sim100$ km/s). We attribute the precession to gravitational interactions with a binary companion after ruling out the presence of a close, dense disk. Under this interpretation, we demonstrate how the morphology of the WR102c helix can be utilized to constrain the orbital period of the system and compare the results with the range of periods expected for interaction between massive stars to occur. Finally, we identify two more massive, evolved stars from previous mid-IR studies (Wachter et al. 2010) that also exhibit possible collimated, helical outflows: the LBV candidate HD316295 (Hillier et al. 1998) and WMD 54, which is broadly characterized as an emission-line B-type star, LBV candidate, or a nitrogen rich WR (WN) star (Wachter et al. 2011). 

\section{Observations}

\subsection{Infrared Imaging}

Observations of WR102c were made using FORCAST (Herter et al. 2013) on the 2.5-m telescope aboard SOFIA. FORCAST is a $256 \times 256$ pixel dual-channel, wide-field mid-infrared camera sensitive from $5 - 40~\mu\mathrm{m}$ with a plate scale of $0.768''$ per pixel and field of view of $3.4'\,\times\,3.2'$. The two channels consist of a short wavelength camera (SWC) operating at $5 - 25~\mu\mathrm{m}$ and a long wavelength camera (LWC) operating at $28 - 40~\mu\mathrm{m}$. An internal dichroic beam-splitter enables simultaneous observation from both long and short wavelength cameras. A series of bandpass filters are used to image at selected wavelengths.

SOFIA/FORCAST observations of the Sickle HII region were taken on the OC1-B Flight 110 on July 2, 2013 (altitude $\sim$ 39,000 ft.) at 19.7, 25.2, 31.5, and 37.1 $\mu\mathrm{m}$. Measurements at 19.7 and 31.5 $\mu$m, as well as 25.2 and 31.5 $\mu$m were observed simultaneously in dual-channel mode, while the 37.1 $\mu$m observations were made in single-channel mode. Chopping and nodding were used to remove the sky and telescope thermal backgrounds. An asymmetric chop pattern was used to place the source on the telescope axis, which eliminates optical aberrations (coma) on the source. The chop throw was $7'$ at a frequency of $\sim 4$ Hz. The off-source chop fields (regions of low mid-infrared Galactic emission) were selected from the Midcourse Space Experiment (MSX) $21~\mu$m image of the Galactic Center. The source was dithered over the focal plane to allow removal of bad pixels and to mitigate response variations. The total integration time was 100 sec at 19.7, 25.2, and 37.1 $\mu$m , and 200 sec at 31.5 $\mu$m. The quality of the images was consistent with near-diffraction-limited imaging: the FWHM of the point spread functions was $3.2''$ at $19.7~\mu$m and $3.8''$ at 37.1 $\mu$m.

Calibration of the images was performed by observing standard stars and applying the resulting calibration factors as described in Herter et al. (2013). Raw data was processed applying the latest techniques for artifact removal and calibration (Herter et al. 2013), and deconvolved to the same final point spread function of $3.8''$. Color correction factors were negligible ($\lesssim5\%$) and were therefore not applied. The 1-$\sigma$ uncertainty in calibration due to photometric error, variation in water vapor overburden, and airmass is $\pm7\%$; however, due to flat field variations ($\sim15\%$), which we are unable to correct for, we conservatively adopt a 1-$\sigma$ uncertainty of $\pm20\%$.

We utilized Paschen-$\alpha$ line (1.87 $\mu$m) and continuum (1.90 $\mu$m) images of the Galactic center (Wang et al. 2010; Dong et al. 2011) taken by the Near Infrared Camera and Multi-Object Spectrometer (NICMOS) on the Hubble Space Telescope. Additionally, we incorporated in our analysis archival mid-IR (5.8 and 8.0 $\mu$m) observations obtained by the Spitzer Space Telescope's Infrared Array Camera (IRAC; Fazio et al. 2004) in the Galactic Legacy Infrared Midplane Survey Extraordinaire (GLIMPSE; Stolovy et al. 2006) as well as archival 24 $\mu$m Multiband infrared Photometer for Spitzer (MIPS, Rieke et al. 2004) observations performed in the MIPSGAL survey (Carey et al. 2009).

Large column densities of dust and gas lead to extreme extinction along lines of sight towards the Galactic center ($A_V \gtrsim 30$). We adopt the extinction curve derived by Fritz et al. (2011) from hydrogen recombination line observations of the minispiral, the HII region in the inner 3 pc of the Galactic center, at 1 - 19 $\mu$m made by the Short Wave Spectrometer (SWS) on the Infrared Space Observatory (ISO) and the Spectrograph for Integral Field Observations in the Near Infrared (SINFONI) on the Very Large Telescope (VLT). We assume a distance towards the Galactic center of 8 kpc (Reid 1993).

\subsection{Infrared Spectroscopy}

Follow-up spectroscopic observations of the WR102c helix (RA = 17:46:12.90, DEC = -28:49:12.5 J2000) were performed on August 6, 2015 using the TripleSpec instrument on the 200'' Hale Telescope at Palomar Observatory. TripleSpec is a medium-resolution ($\lambda / \Delta \lambda\sim2500$) slit spectrograph (Wilson et al. 2004; Herter et al. 2008) that obtains spectra from 1.0 to 2.4 $\mu$m. The TripleSpec slit size is $1\times30''$ and each spectral resolution element is $\sim2.7$ pixels on the $2048\times1024$ pixel array. Spectra were averaged over 6 spatial pixels ($\sim1.8''$) and smoothed by preforming a 3-spectral pixel moving average. 

Due to crowded stellar fields along lines of sight towards the Galactic center and the high background of atomic hydrogen emission lines, $\sim0.3^{\circ}$ amplitude nods were performed to an off-source field of minimal stellar contamination and background emission. Observations were taken in an alternating ABBA nod-sequence with 240 s exposures at each slit position. The total integration time spent on the WR102c helix was 1200 s. The centroids of prominent atmospheric OH emission lines in the off-source nod positions were used to calibrate spectral shifts of the Brackett-$\gamma$ ($\lambda = 2.16 \, \mu$m) along the slit.

\section{Results and Analysis}

\begin{deluxetable*}{cccccccccc}
\tablecaption{Observed Mid- and Far-Infrared Fluxes (in Jy)}
\tablewidth{0pt}
\tablehead{ Region &$\Delta$x & $\Delta$y & $F_{5.8}$ & $F_{8.0}$ &$F_{19}$ &$F_{25}$ &$F_{31}$ &$F_{37}$ &$F_{70}$  }

\startdata
	Helix East & -22.5'' & -7.5'' & - & 0.86 & 17.06 & 20.21 & 18.15 & 17.55 & $\lesssim7$\\
	Helix West & -11.3 & -9.0'' & - & 0.49 & 17.79 & 23.00& 22.02 & 19.97 & $\lesssim7$\\
	Sickle Handle & 6.8'' & -10.5'' & 0.19 & 0.85 & 15.25 & 33.28 & 33.88& 33.28 & $\lesssim31$\\
	Sickle Blade & -112.5'' & 62.5 & 0.33 & 1.23 & 8.52& 22.92& 30.37& 34.73 & $\lesssim40$\\
\enddata

\tablecomments{$\Delta$x and $\Delta$y indicate the angular offset of the region from WR102c (RA: 17:46:11.14 and Dec: -28:49:05.9 J2000, Dong et al. 2011). Square $8.25''\times8.25''$ apertures were used to extract the fluxes. The 1-$\sigma$ errors for the fluxes are assumed to be $20\%$. }
	\label{tab:Flux}
\end{deluxetable*}

\subsection{WR102c Luminosity Correction}

{Barniske, Oskinova, \& Hamann (2008) adopt H and $\mathrm{K}_S$ fluxes of 11.84 and 9.93, respectively, for the photospheric emission from WR102c. However, at 1.9 $\mu$m, which falls in between the H and $\mathrm{K}_S$ bands, the flux from WR102c is reported to be 12.3 mag from observations with HST/NICMOS (Wang et al. 2010). Since the 1.9 $\mu$m flux is not consistent with the brighter H and K band fluxes, we claim that Barniske, Oskinova, \& Hamann (2008) mistakenly adopt the H and $\mathrm{K}_S$ band fluxes from another bright star for their stellar models of WR102c. These models therefore overestimate the stellar luminosity, radius, and mass of WR102c. Since the spectral analysis from Barniske, Oskinova, \& Hamann (2008) is unaffected by the photometric misidentification, its initial classification as a WN6 star remains unchanged (Figer, McLean, \& Morris 1999).

The 2MASS $\mathrm{K}_S$ and H fluxes of WR102c are 11.6 and 13.4, respectively, and were initially reported by Figer, McLean, \& Morris (1999), who studied the population of massive stars in the Quintuplet Cluster and initially identified WR102c. We estimate the stellar luminosity of WR102c from the K-band bolometric correction for WN6 stars provided by Crowther et al. (2006a; $\mathrm{BC}_K\sim3.8$) and the dereddening correction from with the extinction curve derived by Fritz et al. (2011; $A_K=2.5$). A K-band flux of 11.6 mag thus implies a luminosity of $\sim4\times10^5$ $\mathrm{L}_\odot$. This luminosity is consistent with scaling down the flux stellar model of Barniske, Oskinova, \& Hamann (2008) by a factor of $\sim5$, which is consistent with the ratio of the misreported and newly assigned K-band fluxes. Follow-up analysis of WR102c by Steinke et al. (2015) using the proper photometry indeed indicate stellar luminosities of $3 - 4 \times10^5$ $\mathrm{L}_\odot$.

\subsection{Dust and Gas Morphology}

In the mid and far-IR, the helix extends $\sim1.5$ pc from WR102c at the southern base of the Sickle ``handle" (Fig.~\ref{fig:SickleHelix}A) towards the Quintuplet cluster. Although the projected morphology of the handle appears linear, it traces the edge of the dense molecular cloud associated with the Sickle and likely has significant depth along the line of sight. The apparent spatial ``wavelength" of the helix is $\sim1$ pc with a maximum peak-to-trough extent of $\sim0.75$ pc (Fig.~\ref{fig:SickleHelix}B). No other features perpendicular other than the helix are detected along the handle. The inconsistency of the helix with the handle morphology strongly suggests that the helix is not physically linked to the dusty, extended filaments composing the handle, which are shaped by the powerful winds from the Quintuplet cluster and aligned with the magnetic field embedded in the Sickle molecular cloud (Figer, McLean, \& Morris 1999; Dotson et al. 2000; Chuss et al. 2003). We will show, however, that the handle and the helix are close to each other in projection since they both appear to be heated by the radiation field of the Quintuplet cluster. 

There is an identical ionized gas counterpart to the far-IR helix that appears in the image of Paschen-$\alpha$ ($\lambda = 1.87$ $\mu$m) line emission (Wang et al. 2010; Dong et al. 2011; Fig.~\ref{fig:SickleHelix}C). The morphology of the Paschen-$\alpha$ helix is consistent with the structure that appears in the IR, which traces warm dust. At its leading, northwest end the helix coincides with one of two $\sim0.1$ pc-sized lobes that extend from WR102c. Although WR102c itself appears as a Paschen-$\alpha$ line-emitting star, the HeII (8-6, $\lambda=1.8753$ $\mu$m) emission line can reproduce the measured flux (M. Steinke  2015, private communication) since it is claimed to be an H-poor WR star with a stellar temperature of $70000$ K (Steinke et al. 2015). The orientation of these lobes is aligned with the apparent trajectory of the helix near WR102c indicating that they may be linked with the structure of the helix. In the IR, there is a local flux peak at the position of the lobes; however, the emission along lines of sight towards these regions is likely dominated by larger quantities of warm dust in the handle. We note that the helix also appears in the 8.3 GHz continuum map taken by Lang, Goss, \& Wood (1997); however, the signal from the helix is comparable to that of the background flux from which there is also a non-thermal component.

\subsection{Observed Dust Temperature and Heating Sources}
\label{sec:ODT}

We produce a color temperature map using the 19 and 31 $\mu$m images of the Sickle (Fig.~\ref{fig:SickleCT}A) assuming the emission of the dust is optically thin and takes the form $F_\nu  \propto B_\nu  (T_d)\,  \nu^\beta$, where $B_\nu(T_d)$ is the Planck function which depends on the emission frequency, $\nu$, and dust temperature $T_d$, and $\beta$ is the index of the emissivity power-law and assumed to be 2. Although the 19 and 37 $\mu$m images would provide a longer spectral baseline, the signal-to-noise ratio of the extended helix flux detected at 37 $\mu$m is too low to determine a reliable map of color temperatures. Overlaid on the temperature map are the predicted temperature contours for 0.1 $\mu$m-sized (red, dashed) and 0.01 $\mu$m-sized (blue, dotted) silicate grains at the location of the handle heated in equilibrium by the Quintuplet cluster and WR102c. The radiation inputs from the Quintuplet cluster and WR102c are modeled as a 35000 K, $3 \times 10^7$ $L_\odot$ (Figer, McLean, \& Morris 1999) and 35000 K, $4 \times 10^5$ $L_\odot$ (Steinke et al. 2015) point sources, respectively. For 0.1 $\mu$m-sized grains, the observed temperatures at the handle ($T_d \sim 130$ K) far exceed the predicted 85 K and are uniform on projected size scales of $\sim1$ pc in the vicinity of WR102c. Within the helix, the discrepancy between predicted and observed temperatures is even greater: $T_d \sim 95$ K versus 180 K, respectively.  

First, we address the temperature discrepancy at the handle. Although the Quintuplet cluster dominates the heating, WR102c will contribute $\sim25\%$ of the total radiative flux in the surrounding $\sim0.5$ pc vicinity. Despite the inclusion of the heating contribution from WR102c, it is not possible to heat dust in the surrounding $\sim0.5$ pc vicinity of the handle to $\sim130$ K (Fig.~\ref{fig:SickleCT}A) if the dust grains are 0.1 $\mu$m in size. However, given that the handle has been carved out by the powerful Quintuplet cluster winds (Figer, McLean, \& Morris 1999; Simpson et al. 1997), we expect the dust to be kinetically sputtered by the interaction with the gas in the winds (Tielens et al. 1994). Smaller grains exhibit higher temperatures under the same heating conditions as larger grains due to lower heat capacities. By adopting a smaller distribution of grains ($a = 0.01$ $\mu$m) for the handle, the predicted equilibrium temperature $\sim0.5$ pc away from WR102c and $\sim3$ pc from the Quintuplet cluster is $\sim125$ K, which is consistent with the observed temperatures (Fig.~\ref{fig:SickleCT}A). We rule out the possibility that the handle is heated by multiple luminous sources embedded within the Sickle molecular cloud since there are no observed local temperature peaks. 

We now discuss the hotter temperatures observed from the helix ($T_d \sim 180$ K), which are $\sim50$ K higher than the handle and $\sim85$ K higher than the predicted temperatures assuming heating by both the Quintuplet cluster and WR102c. It is difficult to reconcile such high and uniform temperatures, especially across large size scales ($\sim1$ pc). These high and uniform temperatures strongly suggest the helix is composed of transiently heated, very small grains (VSGs) that are smaller than $0.01$ $\mu$m. VSGs will not be heated in equilibrium with the incident radiation field given their small absorption cross-sections and will therefore exhibit much greater temperatures when struck by individual photons (Draine \& Li 2001). Importantly, the spectral shape of the emission from transiently heated VSGs is not sensitive to changes in distance (i.e. incident flux) from its heating source, which is consistent with the temperature uniformity of the helix. In the following section, we perform detailed models of the SEDs from dust throughout the helix and Sickle to substantiate our claims on the dust sizes.  

\begin{figure}[t]
	\centerline{\includegraphics[scale=.2]{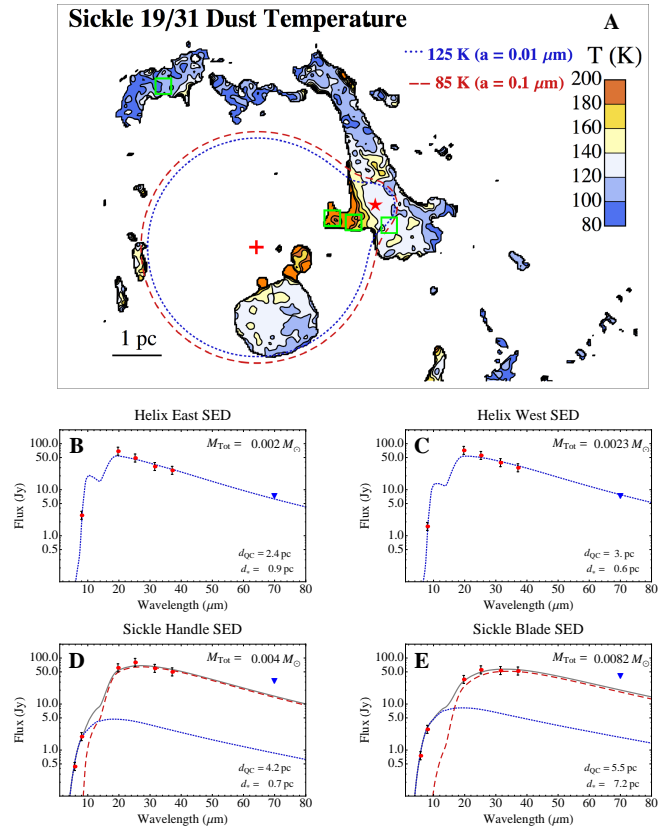}}
	\caption{(A) 19/31 dust temperature map of the Sickle region. Overlaid are the theoretical dust temperature contours intersecting the ``handle" of the Sickle for 0.1 (red, dashed) and 0.01 (blue, dotted) $\mu$m-sized silicate grains. North is up and east is to the left. SEDs are fit to the boxed regions that are the ``Blade'', ``Helix East", ``Helix West", and the ``Handle" from east to west. (B - E) Best-fit DustEM models of the four regions across the helix and Sickle. The blue dotted line in the helix SEDs corresponds to emission from very small grains ($a\sim10\AA$). In the Sickle SEDs the blue dotted line and the red dashed line correspond to emission from very small ($a\sim10\AA$) and small ($a\sim100\AA$) grain size distributions, respectively. The errors of both the Spitzer/IRAC and SOFIA/FORCAST fluxes are assumed to be $20\%$. Due to limits in spatial resolution and fluctuations in the background emission, the measured 70 $\mu$m fluxes from Herschel/PACS (blue triangles) are treated as upper limits.}
	\label{fig:SickleCT}
\end{figure}

\begin{figure*}[t]
	\centerline{\includegraphics[scale=.6]{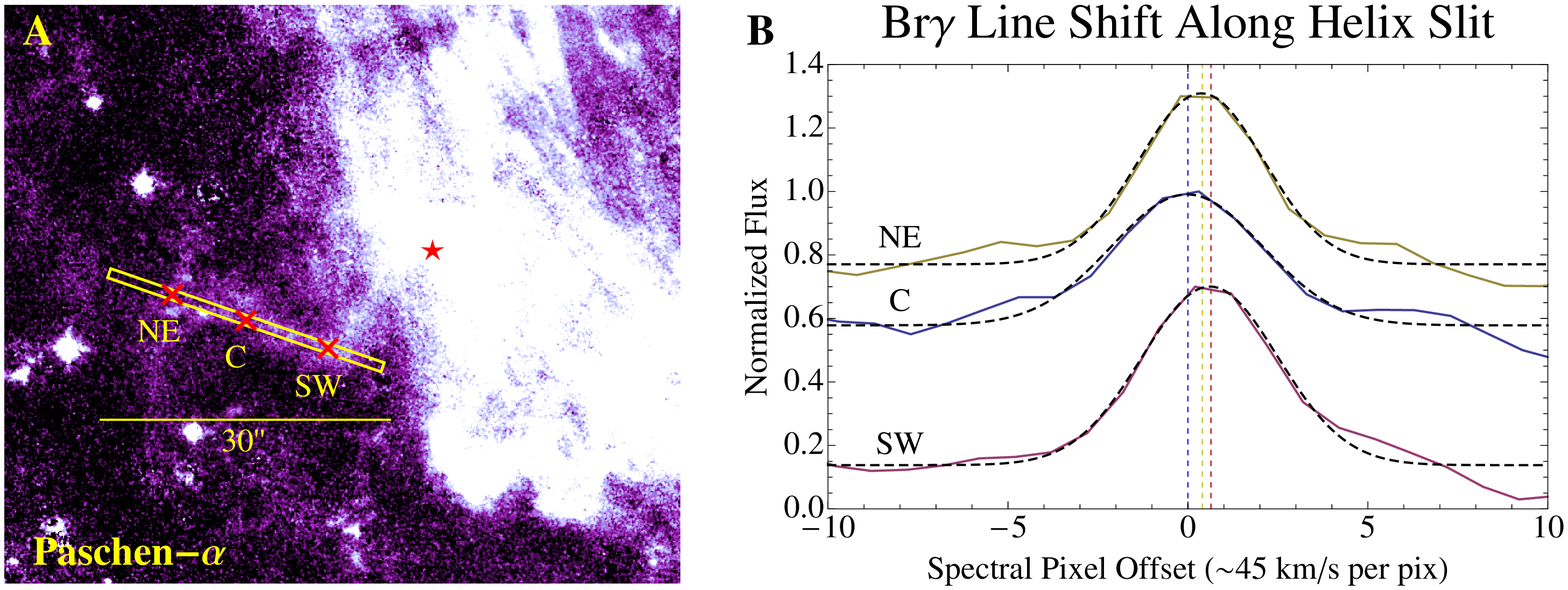}}
	\caption{(A) Paschen-$\alpha$ image of WR102c (red star) and the helix overlaid with the Palomar/TripleSpec slit and the positions where spectra of the Brackett-$\gamma$ line were extracted (red $\times$'s). North is up and east is to the left. (B) Normalized Bracket-$\gamma$ emission line from the NE (yellow), C (blue), and SW (red) positions indicated in (A). The NE and SW lines are offset by $+0.3$ and $-0.3$, respectively. The black dashed lines indicate gaussian fits to the emission line and the colored, dashed vertical lines indicate the spectral pixel offset of each peak with respect to the position C emission line peak. }
	\label{fig:HelixSpectra}
\end{figure*}

\subsection{Dust Spectral Energy Distributions}

Dust models are produced using DustEM (Compi{\`e}gne et al. 2011), which is capable of evaluating spectra of VSG emission. We fit the models to the observed mid- and far-IR SEDs of two regions at the helix (east and west; Fig.~\ref{fig:SickleCT}B and C, respectively) and two regions around the Sickle (the handle and blade; Fig.~\ref{fig:SickleCT}D and E, respectively). The observed fluxes extracted from square $8.25''\times8.25''$ apertures at each region (see Fig.~\ref{fig:SickleCT}A) are listed in Tab~\ref{tab:Flux}. The free parameters of the models are the dust mass abundances and the grain size. The dust is assumed to be heated radiatively by WR102c and the Quintuplet cluster, which are modeled as a $4 \times 10^5 \, L_\odot$ point source (Steinke et al. 2015) with a 35000 K Castelli \& Kurucz (2004) stellar atmosphere and a $3 \times 10^7\, L_\odot$ point source (Figer, McLean, \& Morris 1999) with a 35000 K Castelli-Kurucz (2004) atmosphere, respectively. Separation distances between WR102c and dust in the helix are assumed to be the projected distance divided by cos($\theta_\mathrm{p}$), where $\theta_\mathrm{p}$ is the conical opening angle of the helix ($\sim16^{\circ}$; see Sec.~\ref{Sec:model}). Given the greater uncertainties in the distances between both heating sources and the Sickle and between the Quintuplet Cluster and the helix, a multiplicative factor of $\sqrt{2}$ is included to these projected distances. The 70 $\mu$m flux measurements from the Photoconductor Array Camera and Spectrometer (PACS, Poglitsch et al. 2010) on Herschel (Pilbratt et al. 2010; Molinari et al. 2010) are plotted on the SEDs as a diagnostic for the presence of a cooler distribution of dust and are therefore not included in the model fits.

Model fits show that the dust in the helix can indeed be modeled by a single distribution of transiently heated silicate-type VSGs ($a \sim 10$ \AA), whereas dust in the blade and handle, which trace the edge of the Sickle molecular cloud, are composed of two distributions of small ($a\sim100$ \AA) and very small ($a \sim 10$ \AA) grains. Notably, the 70 $\mu$m flux at the east and west helix regions agrees very closely with the models fit to the mid- and far- IR fluxes and indicates that the helix is not associated with cooler dust in a molecular cloud along the line of sight. This is in direct contrast with SED models of the handle and helix, where the 70 $\mu$m flux excess reveals the presence of cooler dust within the Sickle molecular cloud. The difference in the grain size distributions between the helix and Sickle regions demonstrates that dust composing the helix is independent of the Sickle.

\subsection{Radial Velocity along the Helix}
\label{Sec:vel}

Spectroscopic measurements of the Brackett-$\gamma$ emission line were fit assuming a Gaussian profile at three different spatial positions along the helix (Fig.~\ref{fig:HelixSpectra}A): northeast (NE), center (C), and southwest (SW). The centroid of the emission lines at SW and NE are found to be shifted by $+0.38\pm0.26$ and $+0.62\pm0.28$ spectral pixels relative to the line at C, respectively (Fig.~\ref{fig:HelixSpectra}B). Since the NE line centroid falls within the error of the SW line centroid, we average their relative displacements to the C line centroid and claim that there is a $\sim0.5$ spectral pixel shift between the outer edges of the helix and the central position along the axis of symmetry. A spectral pixel corresponds to $\sim45$ km/s, which implies a relative radial velocity shift of $23\pm12$ km/s between the center and the edges of the helix. The observed velocity shifts along the helix are consistent with the kinematics of a conical outflow, where the motion of gas at the edges is perpendicular to our line of sight and exhibit the highest radial velocity.

\subsection{Gas and Dust Mass}

Electron densities are derived at the SW, C, and NE positions along the helix in order to characterize the density profile and estimate the total ionized gas mass. The densities are estimated from the extinction-corrected (Fritz et al. 2011, Mills et al. 2011) Paschen-$\alpha$ flux assuming case B recombination with an electron temperature identical to that of the Sickle ($T_e\sim5500$ K, Lang, Goss, \& Wood 1997) and an adopted volume based on the spatially resolved thickness of the helix, $t$, at each position. We determine densities of 1.8, 1.6, and 1.4 $\times\,10^3$ $\mathrm{cm}^{-3}$ for positions SW, C, and NE, respectively (Tab.~\ref{tab:HelProp}). Adopting a $20\%$ error in the extinction correction (Fritz et al. 2011) implies that the error of the density estimates is $\pm150$ $\mathrm{cm}^{-3}$. A power-law fit to the densities as a function of projected distance from WR102c, $d_\mathrm{*}$, yields a relatively shallow density gradient with an index of $-0.37\pm0.27$. Assuming a radial density power-law $\propto d_*^{-.37}$ and a uniform thickness of $0.16$ pc throughout the helix, we integrate over the volume of the helix from $d_\mathrm{*}=0.1$ pc, the distance between WR102c and the edge of the lobes (see Fig.~\ref{fig:SickleHelix}C), to $d_\mathrm{*}=1.5$ pc and derive a total ionized gas mass of $\sim0.8$ $\mathrm{M}_\odot$. The total mass of dust composing the helix derived from the best-fit DustEM models is $\sim0.008$ $M_\odot$, which implies a dust-to-gas mass ratio of $\sim1\%$. 

\begin{deluxetable}{ccccc}[h]
\tablecaption{Physical properties at different regions along the helix}
\tablewidth{0pt}
\tablehead{ Region & $d_\mathrm{*}$ (pc)  & $\Delta v$ (km/s) & $t$ (pc) & $n_e$ ($\times\,10^{3}\,\mathrm{cm}^{-3}$) }

\startdata
	SW & $0.6$ & $+28\pm13$ & $0.14\pm0.02$ & $1.8$ \\
	C & $0.8$ & 0 & $0.16\pm0.01$ & $1.6$\\
	NE & $1.1$ & $+17\pm12$ & $0.18\pm0.02$ & $1.4$ \\ 
\enddata

\tablecomments{The electron density, $n_e$, velocity shift of the Brackett-$\gamma$ emission line peak relative to region C, $\Delta v$, and the thickness of the helix, $t$, are provided at different regions along the helix located $d_*$ away from WR102c in projection. The error for the electron densities is $\pm150$ $\mathrm{cm}^{-3}$}
	\label{tab:HelProp}
\end{deluxetable}


\section{Discussion}

\subsection{Precessing Outflow Model of the Helix}
\label{Sec:model}

We interpret the helix as a precessing, highly collimated outflow from WR102c. The morphology of the helix is fit to an analytical model of a precessing, collimated outflow (e.g. Kraus et al. 2006), the apparent shape of which is determined by the outflow velocity, $v_H$, precession period, $\tau_p$, precession angle, $\theta_p$, position angle, $\Theta$, and inclination angle with respect to the plane of the sky, $i$. Based on the Paschen-$\alpha$ morphology of the helix, we adopt $240^\circ$ for $\Theta$, and the outflow velocity is assumed to be $23\pm12$ km/s (see Sec.~\ref{Sec:vel}) divided by $\mathrm{sin}(\theta_p)$. The free parameters are therefore $\theta_p$, $\tau_p$ and $i$, the latter of which we assume is close to $0^\circ$. We fit a precession angle of $16\pm4^\circ$, from which we infer an outflow velocity of $83\pm48$ km/s. The precession period can then be fit with a value of $1.4\pm0.8\times10^4$ yr, which is consistent with the wavelength of the helix ($\sim1$ pc) divided by the outflow velocity (Fig.~\ref{fig:HelixMod}). The model fit implies that the helix has existed for almost one and a half precession periods with a dynamical age of $1.8\pm1.0\times10^4$ yr. The total gas mass of the helix ($\sim0.8$ $M_\odot$) infers a mean mass loss rate of $4.4^{+5.6}_{-1.5} \times 10^{-5}$ $M_\odot$ $\mathrm{yr}^{-1}$. This estimate of the mean mass loss rate should be regarded as a lower limit since the helix only traces a confined, polar outflow from WR102c.

\begin{figure}[t]
	\centerline{\includegraphics[scale=0.42]{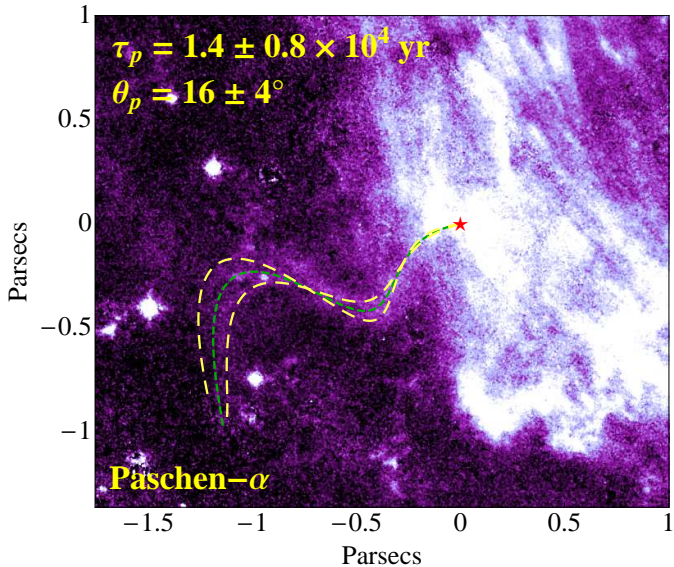}}
	\caption{Precessing outflow model (green dashed line) of the helix extending from WR102c (red star) overlaid on the Paschen-$\alpha$ map of the region. The yellow dashed lines indicate the outflow models with parameters corresponding to the upper and lower error estimates. }
	\label{fig:HelixMod}
\end{figure}

Given the density and velocity of the helix and the outflow properties of the winds from the Quintuplet cluster, we can determine whether or not the morphology of the helix has been influenced by the luminous cluster. We verify that the expansion of the helix is unimpeded by the winds from the Quintuplet cluster by estimating the ram pressure ratio between the Quintuplet winds and the most extended regions of the helix. We define this ram pressure ratio as $\chi_P = P_\mathrm{H}/P_{\mathrm{QC}}$, where

\beq
P_\mathrm{H}=\rho_H v^2_H, P_\mathrm{QC}=\frac{\dot{M}_{QC}}{4\pi \,r_{QC}^2}v_{QC},
\label{eq:pbal}
\eeq

and $\rho_H$ is the density of the helix, $v_H$ is the helix outflow velocity, $\dot{M}_\mathrm{QC}$ is the mass loss rate of the Quintuplet cluster, $r_\mathrm{QC}$ is distance from the Quintuplet, and $v_\mathrm{QC}$ is the velocity of the Quintuplet winds. If $\chi_P>>1$ at regions in the helix most adjacent to the Quintuplet, then we do not expect the Quintuplet winds to have significantly disrupted the morphology of the helix. We assume that the helix extends $\sim1.5$ pc from WR102c (\{-1, -1\} in Fig.~\ref{fig:HelixMod}), which is where $r_\mathrm{QC}\sim2$ pc when including the $\sqrt{2}$ projection factor to account for distance uncertainties between the helix and the Quintuplet. Simulations of mass-loss from the Quintuplet cluster reveal average radial outflow velocities $<400$ km/s at distances of $0.2-1.0'$ ($\sim0.5 - 2.4$ pc) from the cluster (Rockefeller et al. 2005). We therefore adopt an outflow velocity of $v_\mathrm{QC}=400$ km/s. Given the other mass loss and outflow properties of the Quintuplet (Rockefeller et al. 2005; Lang et al. 2005) and the helix, the ram pressure ratio at $r_\mathrm{QC}=2$ pc can be expressed as

\begin{multline}
\chi_P\sim50\left(\frac{\dot{M}_{QC}}{5\times10^{-4}\,M_\odot \,yr^{-1}}\right)^{-1}\left(\frac{v_{QC}}{400\,km\,s^{-1}}\right)^{-1}\\ \left(\frac{d_*}{1.5\,\mathrm{pc}}\right)^{-0.37}\left(\frac{v_H}{80\,km\,s^{-1}}\right)^{2},
\label{eq:pbal2}
\end{multline}

where $d_*$ is the distance from WR102c. As expected, the ram pressure from the helix dominates that of the Quintuplet winds, which indicates that the winds have not yet disrupted the helix. The helix is therefore freely expanding into the cavity of the Sickle and its morphology is consistent with the outflow model (Fig.~\ref{fig:HelixMod}).


\subsection{Emergence of the Helix during a Previous LBV Phase?}
\label{sec:evol}

The disagreement between the observed wind velocity of WR102c ($\sim1600$ km $\mathrm{s}^{-1}$, Steinke et al. 2015) and the outflow velocity of the helix ($\sim80$ km $\mathrm{s}^{-1}$) indicates that the helix did not form in its high velocity winds. Additionally, Steinke et al. (2015) reveal that the bipolar Paschen-$\alpha$ emitting lobes in the $\sim0.15$ pc vicinity of WR102c (Fig.~\ref{fig:SickleHelix}(C)) do not exhibit velocities consistent the winds from WR102c. Discrepant nebular and wind velocities from WR stars are not uncommon since the nebulae are interpreted as ejecta from a previous RSG or LBV phase (Toal\'{a} et al. 2015 and references therein), where the star has undergone enhanced mass loss at speeds of $\sim10-100$ km $\mathrm{s}^{-1}$. Observations of nebulae around H-poor WN-type stars have revealed expansion velocities of $\sim10-100$ km $\mathrm{s}^{-1}$ and enriched metal abundances consistent with ejecta from massive, evolved stars (e.g. S308, RCW58, M1-67, NGC 6888; Chu, Weis, \& Garnett 1999 and references therein). These nebulae have not been significantly accelerated by the more tenuous winds from their central WR stars. Steinke et al. (2015) also suggest that the spectral and morphological characteristics of the bipolar emission around WR102c is consistent with that of a typical WR nebula ejected from a previous phase of enhanced mass loss.

Evolutionary tracks of rotating stars with solar metallicty from Ekstr\"om et al. (2012), which were adopted by Steinke et al. (2015) to derive an initial mass of $\sim40$ $M_\odot$ for WR102c, indicate that stars with an initial mass $\geq30$ $M_\odot$ will not go through an RSG phase (Georgy et al. 2012). We therefore favor the LBV interpretation as the origin of the helix; however, we do not rule out the RSG outflow scenario. The dynamical timescale of the helix ($\sim2\times10^4$ years, Sec.~\ref{Sec:model}) is also comparable to the estimated duration of the LBV phase ($\sim10^4$ years; Garcia-Segura et al. 1996) and supports our claim that WR102c has recently transitioned to a WN star.

\subsection{Evidence for an Unseen, Binary Companion of WR102c}

The precession of WR102c that dictates the morphology of the helix must be provoked by gravitational interactions with either a binary companion or a disk with orbital axes misaligned with the spin axis of WR102c. We rule out the possibility that the precessing outflow is due to interactions with a disk since the strong winds from WR102c would disrupt a nearby disk on very short timescales. The disruption timescale, $\tau_d$, can be estimated as the time it takes a shock driven by winds from WR102c to propagate through the disk. For a disk density and outer radius of $\rho_d$ and $r_d$, respectively, the disruption timescale can be expressed as

\beq
\tau_d = \sqrt{\frac{\rho_d}{\rho_w}} r_d/v_w,
\label{eq:disrupt1}
\eeq

where $\rho_w$ is the density of the winds from WR102c and $v_w$ is the wind velocity. In order to estimate $\tau_d$, we adopt a gas density of $\rho_d\sim10^{-10}$ g $\mathrm{cm}^{-3}$ and radius of $r_d\sim200$ $\mathrm{R}_\odot$, which are consistent with near-IR observations of disks inferred around rapidly rotating B-type (Be) stars (Gies et al. 2007). Assuming the density of the medium surrounding the disk is dominated by the mass loss, $\dot{M}_w$, and winds from WR102c (i.e. $\rho_w\approx\frac{\dot{M}_{w}}{4\pi \,r_{d}^2 v_w}$), the disruption timescale is

\begin{multline}
\tau_d\sim1.8\left(\frac{\dot{M}_{w}}{4\times10^{-5}\,M_\odot \,yr^{-1}}\right)^{-1/2}\left(\frac{v_{w}}{80\,km\,s^{-1}}\right)^{-1/2}\\ \left(\frac{r_d}{200\,\mathrm{R}_\odot}\right)^{2}\left(\frac{\rho_d}{ 10^{-10} \mathrm{g} \, \mathrm{cm}^{-3}}\right)^{1/2}\,\mathrm{yr},
\label{eq:disrupt2}
\end{multline}

where we have adopted the mass loss rate derived in Sec.~\ref{Sec:model} and a wind velocity equivalent the estimated helix outflow speed. Even if the disk were several orders of magnitude denser, the disruption timescale due to the shocks driven by WR102c would still be shorter than the inferred lifetime of the helix, $\sim1.8\times10^4$ yr. We also note that it is unlikely a disk formed due to enhanced equatorial mass-loss from WR102c via rotational distortion since the resulting disk would be aligned with the star's rotational axis and thus provide no torque for the star to precess. Additionally, equatorial mass-loss is predicted to be considerably suppressed for massive, oblate stars due to gravity darkening at the equator and pole-ward latitudinal line-driven forces (Owocki, Cranmer, \& Gayley 1996). 

We therefore conclude that the perturber must be a binary companion. The binarity of WR102c not only explains its precession, but also provides a plausible scenario for the ejection and subsequent isolation from its likely birth site, the Quintuplet cluster, as well as a rapid spin-up of WR102c that can produce an enhanced polar outflow (Dwarkadas \& Owocki 2002). 

We hypothesize that WR102c and an initially more massive binary companion were formed and subsequently ejected from the Quintuplet cluster. The association of WR102c with the Quintuplet is substantiated by the presence of similarly evolved, massive stars within the cluster (Liermann, Hamann, \& Oskinova 2009). Based on the gainer/donor scenario for massive stars in interacting, mass-exchanging binaries (e.g. Smith \& Tombleson 2015), we interpret the progenitor of WR102c to be the mass gainer of the system and its companion the mass donor. The donor, which is the star with the greatest initial mass in the system, evolves faster than the gainer and begins to fill its Roche lobe as its envelope swells while leaving the main sequence. Mass from the envelope of the donor accretes on to the gainer due to Roche lobe overflow and also spins up the gainer to near Keplerian, break-up speed (e.g. de Mink et al. 2013). The donor then explodes as a supernova (SN), and assuming its explosion is spherically symmetric, the gainer and remnant core of the donor will still remain bound since the donor lost a significant fraction of its initial mass to winds and mass transfer. The SN explosion will, however, provide a velocity kick to the center of mass of the bound system and eject it from its birth site  (e.g. Blaauw 1961). Based on population synthesis models of massive stellar binaries, $\sim20\%$ of binaries are expected to remain bound after the initial SN (Eldridge, Langer, \& Tout 2011). 

We do not rule out the possibility that the WR102c system was ejected via dynamical interactions with other members of the Quintuplet. However, it is unlikely that the companion is a closely orbiting MS or post-MS star given the lack of a detectable X-ray counterpart within a $\sim30''$ radius of WR102c that typically arise from wind collision zones between two massive stars (Mauerhan et al. 2009). We also rule out the possibility that the outflow originates from an accretion disk around a compact companion due to the lack of a hard X-ray counterpart and the hostile conditions for dust surival in a high velocity jet-like outflow from a compact object.

\subsection{Precession in a WR102c Binary System}

\begin{figure}[t]
	\centerline{\includegraphics[scale=0.5]{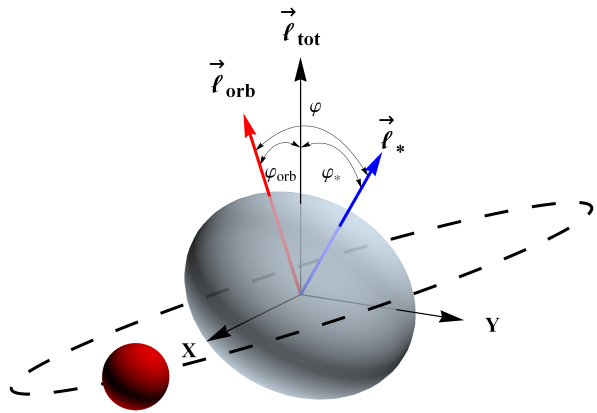}}
	\caption{Illustration of the precession geometry in the WR102c system. The z-axis is aligned with the total angular momentum vector of the system, $\vec{\ell}_\mathrm{tot}=\vec{\ell}_\mathrm{*}+\vec{\ell}_\mathrm{orb}$. The grey ellipsoid and red sphere correspond to WR102c and its companion, respectively. The blue arrow indicates the direction of the spin angular momentum vector of WR102c, $\vec{\ell}_\mathrm{*}$, and the red arrow indicates the direction of the orbital angular momentum vector of the binary companion, $\vec{\ell}_\mathrm{orb}$. Objects are not drawn to scale.}
	\label{fig:Prec}
\end{figure}

The binary companion will exert a torque on WR102c that attempts to align its spin axis with the angular momentum vector of the orbital plane, $\vec{\ell}_\mathrm{orb}$. A torque is also applied on $\vec{\ell}_\mathrm{orb}$ from the misaligned spin angular momentum vector of WR102c, $\vec{\ell}_\mathrm{*}$. The result is a mutual precession of both spin and orbital axes about the net angular momentum vector of the system, $\vec{\ell}_\mathrm{tot}=\vec{\ell}_\mathrm{orb}+\vec{\ell}_\mathrm{*}$. This precession scenario is analogous to planetary systems with a hot Jupiter in a misaligned orbit with its host star's spin axis (e.g. Barnes et al. 2013). Figure~\ref{fig:Prec} illustrates the precession geometry of the WR102c system where the z-axis is aligned with $\vec{\ell}_\mathrm{tot}$, $\varphi_*$ is the angle between $\vec{\ell}_\mathrm{*}$ and $\vec{\ell}_\mathrm{tot}$, $\varphi_\mathrm{orb}$ is the angle between $\vec{\ell}_\mathrm{orb}$ and $\vec{\ell}_\mathrm{tot}$, and $\varphi=\varphi_*+\varphi_\mathrm{orb}$. The angles and amplitudes of the angular momentum from WR102c's spin and the orbital motion of the companion can be related by 

\beq
\ell_\mathrm{*}\,\mathrm{sin}(\varphi_*)=\ell_\mathrm{orb}\,\mathrm{sin}(\varphi_\mathrm{orb}).
\label{eq:LL}
\eeq

In the case where $\vec{\ell}_\mathrm{orb}>>\vec{\ell}_\mathrm{*}$ with low orbital eccentricity and $\varphi<<90^\circ$, $\vec{\ell}_\mathrm{orb}\approx\vec{\ell}_\mathrm{tot}$ and the spin axis of WR102c precesses about the angular momentum vector of the orbital plane at a frequency of
\beq
\Omega_*\equiv\frac{2\pi}{\tau_p}=\frac{3 \pi^2}{\omega_* P^2}\left(\frac{M_B}{M_A+M_B}\right)\left(1-\left(\frac{R_p}{R_{eq}}\right)^2\right)\mathrm{cos}(\varphi),
\label{eq:prec}
\eeq

where $\tau_p$ is the precession period, $P$ is the orbital period of the system, $M_A$ is the mass of WR102c, $M_B$ is the mass of the binary companion, $\omega_*$ is the rotational frequency of WR102c, and $R_p$ and $R_{eq}$ are the polar and equatorial radii of WR102c. In the case where $\vec{\ell}_\mathrm{orb}\sim\vec{\ell}_\mathrm{*}$, the precession rate of WR102c is faster since the angle $\varphi_\mathrm{*}$ between $\vec{\ell}_\mathrm{*}$ and $\vec{\ell}_\mathrm{tot}$ is now smaller than the angle $\varphi$ between $\vec{\ell}_\mathrm{*}$ and $\vec{\ell}_\mathrm{orb}$. The precession rate of WR102c in this more general case can be expressed as

\beq
\Omega'_*\equiv \frac{2\pi}{\tau'_p}=\Omega_* \frac{\mathrm{sin}(\varphi)}{\mathrm{sin}(\varphi_*)},
\label{eq:OC2}
\eeq

where $\tau_p'$ is the precession period of WR102c about the net angular momentum vector. It can be shown that $\ell_\mathrm{orb}\gtrsim\ell_\mathrm{*}$ for the WR102c system, which implies $\varphi_*\gtrsim\varphi_\mathrm{orb}$ (Eq.~\ref{eq:LL}) and $\varphi\lesssim 2\varphi_*$. Since the observed half-opening angle of the helix, $\theta_p$, is the precession angle of WR102c, $\varphi_*$, it follows that $\varphi\lesssim 2\theta_p\sim32^\circ$ (Sec.~\ref{Sec:model}). We may therefore approximate $\mathrm{sin}(\varphi)$, $\mathrm{sin}(\varphi_*)$, and $\mathrm{sin}(\varphi_\mathrm{orb})$ as $\varphi$, $\varphi_*$, and $\varphi_\mathrm{orb}$, respectively. Equation~\ref{eq:OC2} can then be re-expressed as 

\beq
\Omega'_*\approx\Omega_* \frac{\varphi}{\varphi_*}=\Omega_*\left(1+\frac{\varphi_\mathrm{orb}}{\varphi_*}\right)\approx\Omega_*\left(1+\frac{\ell_*}{\ell_\mathrm{orb}}\right),
\label{eq:OC3}
\eeq

where we have also applied Eq.~\ref{eq:LL}. The magnitude of the spin angular momentum of WR102c is given by ${\ell}_\mathrm{*}=k\,M_A\,R_{eq}^2\,\omega_*$, where $k$ is the moment of inertia coefficient (e.g. $k=0.4$ for a uniform density sphere and $k\sim0.06$ for the Sun). For the binary companion with an orbital radius, $a$, and orbital frequency, $n_B\equiv\frac{2\pi}{P}$, the magnitude of the orbital angular momentum is simply $\ell_\mathrm{orb}=M_B\,a^2\,n_B$. The upper limit of $\frac{\ell_*}{\ell_\mathrm{orb}}$ can be obtained by assuming the physical limits where $a\sim R_{eq}$ and that WR102c is rotating near the Keplerian breakup velocity, $\omega_* \sim \sqrt{\frac{G\,M_A}{R_{eq}^3}}$. Given our scenario from the previous section, the companion is interpreted as the remnant core of a massive star which will exhibit a mass no less then $M_B\sim2$ $\mathrm{M}_\odot$. Recall that Steinke et al. (2015) estimate an initial mass of $\sim40$ $M_\odot$ for WR102c based on evolutionary tracks for rotating stars from Ekstr\"om et al. (2012). In the limit where $M_B<<M_A$ and $a\sim R_{eq}$, we assume the system is tidally locked and that $n_B\sim\omega_*$.  The moment of inertia coefficient is estimated to be $k\sim0.05$ from formulae fit to evolutionary models of $k$ by Pols et al. (1998) as a function of initial stellar mass and radius (see Sec. A.2.1 in de Mink et al. 2013). Adopting these limiting physical properties of the WR102c system, the upper limit of $\frac{\ell_\mathrm{*}}{\ell_\mathrm{orb}}$ is

\beq
\frac{\ell_\mathrm{*}}{\ell_\mathrm{orb}}=\frac{k\,M_A\,R_{eq}^2\,\omega_*}{M_B\,a^2\,n_B} \lesssim 1 \left(\frac{k}{0.05}\right)\left(\frac{M_{A}}{40\,\mathrm{M}_\odot}\right)\left(\frac{M_{B}}{2\,\mathrm{M}_\odot}\right)^{-1},
\label{eq:limit}
\eeq

which shows that $\ell_\mathrm{orb}\gtrsim\ell_\mathrm{*}$. Equations~\ref{eq:prec} and~\ref{eq:OC3} can then be combined to redefine $\Omega'_*$ as

\begin{multline}
\Omega'_*\approx\frac{3 \pi^2}{\omega_* P^2}\left(\frac{M_B}{M_A+M_B}\right)\left(1-\left(\frac{R_p}{R_{eq}}\right)^2\right)\mathrm{cos}(\varphi) \\\left(1+\frac{k\,M_A\,R_{eq}^2\,\omega_*}{M_B\,\left(P/2\pi\right)^{1/3}\,(G(M_A+M_B))^{2/3}}\right),
\label{eq:prectot}
\end{multline}

where we have also utilized Kepler's 3rd law: $n^2_B\,a^3=G(M_A+M_B)$. With the observed and adopted parameters of the WR102c system, we can utilize Eq.~\ref{eq:prectot} to constrain the orbital period of the binary system. Since we claim the helix was formed during a previous LBV phase of WR102c (Sec.~\ref{sec:evol}), the stellar radius inferred from recent observations ($\sim4$ $\mathrm{R}_\odot$; Steinke et al. 2015) is unlikely consistent with its radius while an LBV. We therefore adopt the measured radius and rotational velocity for AG Carinae (Groh, Hillier, \& Damineli 2006), an LBV that exhibits rapid rotation, a bipolar nebula, and a dusty helical ``jet" (Paresce \& Nota 1989, Nota et al. 1992, 1995). The radius and spin of AG Carinae averaged over three observed epochs are $R_{eq}=82$ $\mathrm{R}_\odot$ and $\omega_* =0.68\,\omega_{K}$, respectively, where $\omega_K$ is the Keplerian breakup velocity and the inclination of the star is assumed to be $90^\circ$ (Groh, Hillier, \& Damineli 2006). Optical observations show that fast rotation is typical for all bona fide, strong variable galactic LBVs (Groh et al. 2009), thus a rapidly rotating LBV progenitor for WR102c is not unexpected. Binary population synthesis models performed by de Mink et al. (2013) also predict that $\sim20\%$ of all massive main sequence stars are rapid rotators due to binary interaction, which is consistent with the observed fraction of rapid rotators (e.g. Penny \& Gies 2009).

The remaining parameters required to constrain $P$ from Eq.~\ref{eq:prectot}, are $R_{eq}/R_p$, $\varphi$, $\Omega_*'$, $M_A$, and $M_B$. The oblateness of the star, $R_{eq}/R_p$, can be related to its spin (see A.2.2 of de Mink et al. 2013). Since $\theta_p<\varphi\lesssim 2\theta_p$ and $\theta_p$ is a relatively small angle ($\sim16^\circ$), it follows that $\mathrm{cos}(\varphi)\sim\mathrm{cos}(\theta_p)$. The stellar precession rate of WR102c, $\Omega'_*$, is simply $2\pi/\tau'_p$, where $\tau'_p$ is the measured precession period of the helix ($\sim1.4\times10^4$ yr, Sec.~\ref{Sec:model}). For the mass of WR102c during the production of the helix, $M_A$, we adopt the estimated initial mass value of $40$ $\mathrm{M}_\odot$ estimated by Steinke et al. (2015). The orbital period can then be constrained by assuming the range of masses exhibited by compact, remnant cores from core-collapse SNe (neutron stars and black holes) for $M_B$. Since current observations of stellar mass black holes in the Milky Way exhibit a narrow mass distribution at $7.8\pm1.2$ $\mathrm{M}_\odot$ (\"{O}zel et al. 2010 and ref. therein), we use 2 and 8 $M_\odot$ as the lower and upper mass limit for $M_B$, respectively. We determine from Eq.~\ref{eq:prectot} that the orbital period of the WR102c binary system is constrained to $800\,\mathrm{d}< P<1400$ d. The limits derived for $P$ are consistent with the dynamical regime for a relatively wider orbit for massive binary systems where accretion and spin up are predicted occur without a merger event ($2\,\mathrm{d}\lesssim P\lesssim1500$ d; de Mink et al. 2013; Sana et al. 2012; Podsiadlowski, Joss, \& Hsu 1992). Orbital and outflow properties of the WR102c system are summarized in Tab.~\ref{tab:PrecProp}.

\begin{deluxetable}{ccccc}[b]
\tablewidth{0pt}
\tablecaption{Derived helix outflow and orbital properties of the WR102c system}
\tablehead{ $v_H$ (km/s) & $\dot{M}$ & $\theta_p$ $(^\circ)$ & $\tau_p$ ($10^4$ yr) & $P$ (days) }

\startdata
	$80\pm43$ & $4.4^{+5.6}_{-1.5}$  & $16\pm4$ & $1.4\pm0.8$ & $800<P<1400$\\ 
\enddata

\tablecomments{The helix outflow velocity, $v_H$, is derived from the velocity shift of the Brackett-$\gamma$ emission line peak along the helix, and the mass loss rate, $\dot{M}$ (in $10^{-5}\,M_\odot$ $\mathrm{yr}^{-1}$), is the total observed gas mass in the helix divided by its dynamical timescale, $1.8\times10^4$ yr. The precession period of WR102c, $\tau_p$, and the precession angle, $\theta_p$, are derived from an analytical precessing outflow model. $P$ is the orbital period of the system constrained from Eq.~\ref{eq:prectot}. }
	\label{tab:PrecProp}
\end{deluxetable}

\subsection{Helices around Other Massive Evolved Stars}

\begin{figure*}[t]
	\centerline{\includegraphics[scale=0.5]{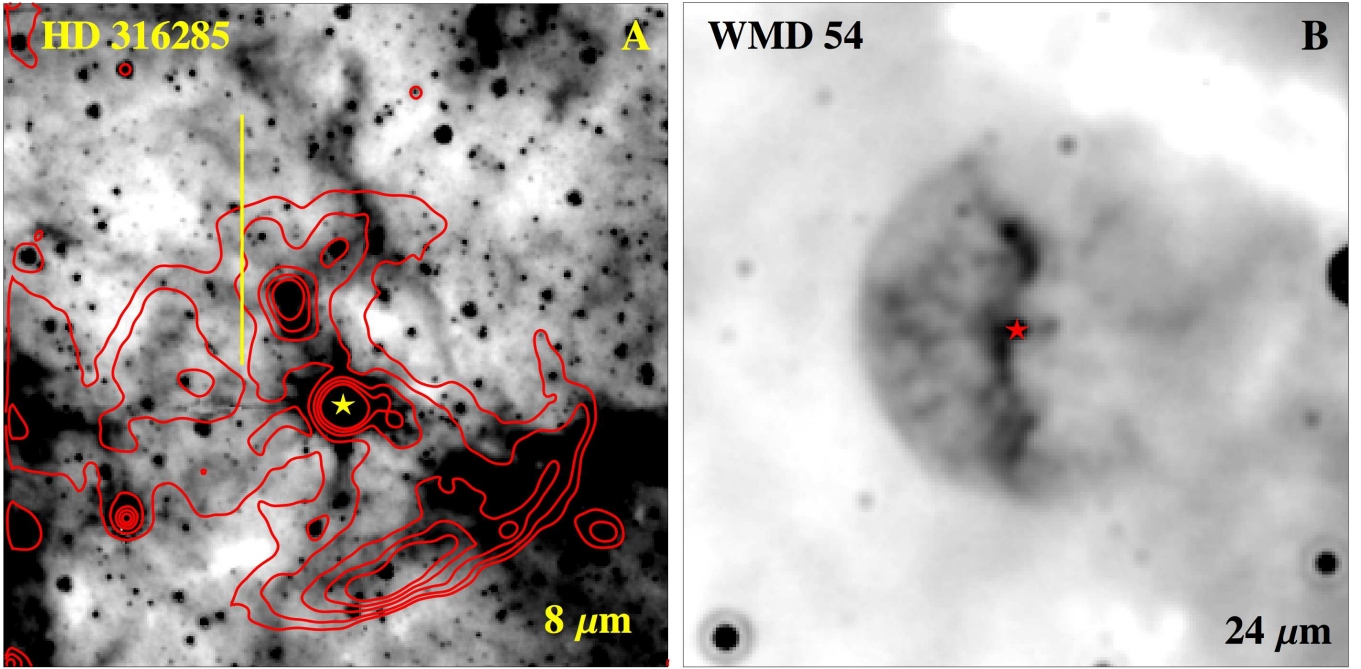}}
	\caption{(A) 8 $\mu$m image of a helix extending from HD 316285 (yellow star) taken by Spitzer/IRAC with linear 24 $\mu$m contours overlaid in red. The yellow vertical line indicates the length of $1$ pc assuming $d=2$ kpc.  (B) 24 $\mu$m image of WMD 54 (red star) and the surrounding helix/nebula taken by Spitzer/MIPS. North is up and east is to the left.}
	\label{fig:HelixIm}
\end{figure*}

The presence of a helix due to a precessing, collimated outflow is unlikely unique to WR102c given the prevalence of binarity in massive stars (Sana et al. 2012). However, we expect helical trails from massive binary systems to be relatively rare since they require a bound system after the more massive companion explodes as a supernova (Eldridge, Langer, \& Tout 2011) and inclination angles favorable for observations. We investigated $\sim60$ nebulae surrounding massive evolved stars identified by Spitzer/MIPS imaging observations at 24 $\mu$m (Wachter et al. 2010) and found two sources that exhibit helical morphologies. The stars consistent with the helices are HD 316285 (Fig.~\ref{fig:HelixIm}a; Hillier et al. 1998) and the source number 54 (WMD 54; Fig.~\ref{fig:HelixIm}b) identified by Wachter et al. (2010).

HD 316285 is classified as a candidate luminous blue variable (LBV) based on its spectral variability and is believed to be located $\sim2$ kpc away (Hillier et al. 1998). The 24 $\mu$m nebular emission surrounding HD 316285 exhibits a bipolar morphology consistent with the orientation of the apparent helix detected at 8 $\mu$m (Fig.~\ref{fig:HelixIm}A), which resembles the bipolar Paschen-$\alpha$ lobes and helix of WR102c (Fig.~\ref{fig:SickleHelix}C). Proper motion measurements of HD 316285 indicate the star is moving south-southwest at a rate of $\sim5$ mas/yr (van Leeuwen 2007). If the nebula is a product of the interaction of an outflow from HD 316285 with the surrounding medium, the enhanced emission at its southern edge at 24 $\mu$m is consistent with HD 316285's direction of motion. This may explain the one-sided appearance of the helix in HD 316285, which is similar to the scenario we infer for WR102c. Additionally, the proper motion velocity of HD 316285 is $\sim45$ km/s assuming a distance of 2 kpc and is consistent with velocities exhibited by systems that remained bound after the initial supernova kick (van den Heuvel et al. 2000). The size of the spatial wavelength of the helix is $\sim0.8$ pc and its total length is $\sim1.6$ pc assuming the HD 316285 is located 2 kpc away. If the helix is indeed associated with a precessing outflow from HD 316285, we infer a precession period of $\tau_p\sim1900$ yr and a total dynamic lifetime of $\sim3800$ yr given the measured wind velocity of $\sim410$ km/s (Hillier et al. 1998). The lifetime of the helix is therefore consistent with HD 316285 being an LBV since it is less than the estimated timescale for the LBV phase ($\sim10^4$ yr). By adopting $\tau_p\sim1900$ yr for Eq.~\ref{eq:prectot} and assuming the same physical parameters as we did for WR102c, the orbital period of a hypothetical HD 316285 binary system may be constrained to $300\,\mathrm{d}< P<520$ d, which agrees with the range of orbital periods where massive binaries are expected to interact.

WMD 54 is broadly characterized as an emission-line B-type star, LBV candidate, or late-type nitrogen-rich Wolf-Rayet star based on the presence of both hydrogen and helium emission lines (Wachter et al. 2011). Unlike WR102c and HD 316285, the apparent helix at 24 $\mu$m around WMD 54 exhibits a symmetric morphology (Fig.~\ref{fig:HelixIm}B). The symmetric appearance of the helix may be consistent with the dynamics of the system if the enhanced emission at the eastern edge of the nebula is indicative of an eastward proper motion of the star. However, neither the distance to WMD 54 nor its proper motion is well-constrained since the system is not well-studied. Preliminary analysis of near-IR spectra of the WMD 54 helix taken recently by Palomar/Triplespec do not reveal the presence of Br-$\gamma$ emission lines, and we are therefore unable to infer its kinematics. The lack of Br-$\gamma$ emission, which suggests the radiation field from WMD 54 is softer than typical WR stars, and the absence of broad $\sim1000$ km/s helium emission lines (Wachter et al. 2011) indicates it that WMD 54 is not likely a WR star. An LBV classification of WMD 54 would strengthen our claim of the formation of  dusty helices during the LBV phase if the helical feature is indeed associated with an outflow from the central star. Further observations are required to verify this hypothesis.

\section{Conclusions}

We presented evidence of a dusty helical outflow extending from the massive, evolved star WR102c from mid- to far-IR imaging observations of the Sickle HII region taken by SOFIA/FORCAST. Based on the following analyses, we determined that the helix is not associated with the dust and gas composing the Sickle, but is instead consistent with a highly collimated outflow from WR102c: 1) The helix extends in a direction perpendicular to the handle of the Sickle region, which is inconsistent with the coherent and ``sheared" filamentary morphology of the handle (Fig.~\ref{fig:SickleHelix}A). 2) Color temperatures of dust composing the helix are much hotter than that of the Sickle and are inconsistent with being equilibrium-heated by the radiation from WR102c and the Quintuplet cluster (Fig.~\ref{fig:SickleCT}A). Models fit to the SEDs of dust in the helix reveal that it is composed of transiently-heated, very small ($\sim10\, \AA$) grains  (Fig.~\ref{fig:SickleCT}B- E). The formation of very small grains in the helix is consistent with condensing in the outflow from a massive, evolved star. 3) There is a lack of cool dust emission associated with the helix, which suggests it is not tracing an illuminated edge of a molecular cloud like the Sickle.

The helix is interpreted as a precessing, highly collimated outflow from WR102c, and we claim it was formed during a previous LBV or RSG phase based the observed $\sim80$ km/s outflow velocities and the nature of nebulae observed around other WR stars (Toal\'{a} et al. 2015 and ref. therein). We favor the LBV interpretation given the absence of an RSG phase in the Ekstr\"om et al. (2012) evolutionary track for a rotating star with an initial mass of $\sim40$ $M_\odot$ (Georgy et al. 2012; Steinke et al. 2015). We however do not rule out the possibility that the helix formed during an RSG phase. We consider two possible mechanisms for inducing the precession: a binary companion or disk. Since it is unlikely that a disk would survive in the vicinity of WR102c, we claim that the precession is induced by an unseen binary. Additionally, there is no hard X-ray counterpart to WR102c that would indicate wind collision regions (Mauerhan et al. 2009), which suggests that the companion is not a main-sequence or post-MS star. We fit precessing outflow models to the morphology of the helix (Fig.~\ref{fig:HelixMod}) and are able to constrain the orbital period of the WR102c system to $800\,\mathrm{d}<P<1400$ d from physical limits we apply to the system and from adopting stellar parameters of the rapidly rotating LBV with a bipolar nebula and dusty helical ``jet", AG Carinae (Groh, Hillier, \& Damineli 2006). Our estimate is consistent with the range of massive binary orbital periods where spin-up and interaction via mass exchange is expected to occur ($P\lesssim1500$ d; de Mink et al. 2013; Sana et al. 2012; Podsiadlowski, Joss, \& Hsu 1992). 

Although helical outflows from massive, evolved stars are likely to be rare, other helices should be detectable from previous imaging observations. A search through the mid-IR observations of nebulae surrounding massive, evolved stars (Wachter et al. 2010) revealed two sources that exhibit helices and warrant further study (Fig.~\ref{fig:HelixIm}): the LBV candidate HD 316295 (Hillier et al. 1998) and WMD 54, which exhibits emission lines consistent with an emission-line B-type star, LBV candidate, and a nitrogen-rich WR star (Wachter et al. 2011). If the helix from HD 316285 is produced under similar circumstances to the WR102c helix, the orbital period of a hypothetical HD 316285 binary system can be roughly constrained to $300\,\mathrm{d}< P<520$ d, which also agrees with the range of orbital periods where massive binaries are expected to interact. Verification of HD 316285 and WMD 54 as bona fide LBVs would strengthen our claim that the formation of dusty helices are a phenomena associated with the LBV phase. High spectral and spatial resolution velocity maps of the helices will be required to confirm their origin as highly collimated outflows from massive evolved stars. There are several important questions regarding the nature of the helices that are beyond the scope of this study are: e.g. what would be the collimation mechanism of the outflow, and how would the helical morphology evolve with the changing radii and rotation rate of the central stars? 

Confirmation of the helices as outflows would provide another observational avenue to study the influence of binarity on the mass loss from massive, evolved stars. The identification of these helical structures could provide a new observational approach to probe the orbital dynamics of massive binary systems. Since most massive stars occur in close, interacting binaries (Sana et al. 2012), these studies can reveal important dynamical properties of immediate progenitors to supernovae and gamma ray bursts. 

\textit{Acknowledgments}. We would like to thank the rest of the FORCAST team, Joe Adams, George Gull, Justin Schoenwald, and Chuck Henderson, the USRA Science and Mission Ops teams, and the entire SOFIA staff. R. L. would like to thank Dong Lai, Selma de Mink, Nathan Smith, and the anonymous referee for the valuable feedback and discussion on binaries and massive stars. R. L. would also like to thank Martin Steinke and Lida Oskinova for the insightful exchanges on WR102c. This work is based on observations made with the NASA/DLR Stratospheric Observatory for Infrared Astronomy (SOFIA) and upon work supported by the National Science Foundation Graduate Research Fellowship under Grant No. DGE-1144153. This work is also based in part on observations obtained at the Hale Telescope, Palomar Observatory as part of a continuing collaboration between the California Institute of Technology, NASA/JPL, Oxford University, Yale University, and the National Astronomical Observatories of China as well as work in part on archival data obtained with the Spitzer Space Telescope, which is operated by the Jet Propulsion Laboratory, California Institute of Technology under a contract with NASA. A portion of this work was carried out at the Jet Propulsion Laboratory, California Institute of Technology, under a contract with the National Aeronautics and Space Administration. SOFIA science mission operations are conducted jointly by the Universities Space Research Association, Inc. (USRA), under NASA contract NAS2-97001, and the Deutsches SOFIA Institut (DSI) under DLR contract 50 OK 0901. Financial support for FORCAST was provided by NASA through award 8500-98-014 issued by USRA.

\end{document}